\documentclass[12pt]{article}
\usepackage[numbers,sort]{natbib}
\usepackage{epsfig, amssymb, amsmath, multicol, setspace}

\usepackage{geometry}
\usepackage[layout=modern]{advancedcoverpage}

\newcommand{\fb}[1]{
  \begin{center}
    \fbox{\parbox{.85\textwidth}{#1}}
  \end{center}
}
\title{In Situ Probes of the First Galaxies and Reionization: Gamma-ray Bursts}
\subtitle{A Whitepaper Submitted to the Decadal Survey Committee}
\presentedto{\underline{Science Frontier Panels:}\\{\it PRIMARY: Cosmology and Fundamental Physics (CFP)} \\ {\it SECONDARY: Stars and Stellar Evolution (SSE), Galaxies across Cosmic Time (GCT)}} 
\author{\emph{\underline{Authors}}\\Matthew McQuinn (Harvard/CFA,  mmcquinn@cfa.harvard.edu, phone: 617-495-2536), Joshua S. Bloom (UC Berkeley), Jonathan Grindlay (Harvard/CFA), 
David Band (GSFC),
S.D. Barthelmy (GSFC),
E. Berger (Harvard-CfA),
A. Corsi (INAF-Roma),
S. Covino (INAF-Brera),
G.J. Fishman  (MSFC),
Steven R. Furlanetto (UCLA),
Neil Gehrels  (GSFC),
D.H. Hartmann  (Clemson),
Chryssa Kouveliotou  (MSFC),
A.S. Kutyrev (GSFC),
Abraham Loeb  (Harvard/CfA),
S. Harvey Moseley  (GSFC),
Tsvi Piran   (Racah Inst.),
L. Piro    (INAF-Rome),
J.X. Prochaska (UCSC), 
R. Salvaterra  (INAF-Brera),
P. Schady (University College London-MSSL),
 A. M. Soderberg  (Harvard-CFA),
 G. Tagliaferri  (INAF-Brera)
}
\organization{\underline{Projects/Programs Emphasized:}\\The Energetic X-ray Imaging Survey Telescope (EXIST); http://exist.gsfc.nasa.gov\\
James Webb Space Telescope/Thirty Meter Telescope/Giant Magellan Telescope}
\date{}

\begin{document}
\maketitle

\def\ale{\mathrel{\hbox{\rlap{\hbox{\lower4pt\hbox{$\sim$}}}\hbox{$<$}}}}
\def\age{\mathrel{\hbox{\rlap{\hbox{\lower4pt\hbox{$\sim$}}}\hbox{$>$}}}}

\newcommand\ion[2]{#1$\;${\small\rmfamily\sc{#2}}\relax}%

\newcommand{\avgchi}{\bar{\chi}_{\rm H I}}
\newcommand{\barxi}{{\bar{x}_i}}
\newcommand{\Mpc}{{\rm Mpc}}
\newcommand{\MHz}{{\rm MHz}}
\newcommand{\Msun}{{M_{\odot}}}
\newcommand{\bfM}{{\boldsymbol{M}}}
\newcommand{\bfP}{{\boldsymbol{P}}}
\def\apj{ApJ}
\def\apjl{ApJL}
\def\apjs{ApJS}
\def\aj{AJ}
\def\mnras{MNRAS}
\def\physrep{physrep}
\def\pasj{PASJ}
\def\araa{{Ann.\ Rev.\ Astron.\& Astrophys.\ }}
\def\aap{{\em A.\&A}}
\def\prd{PRD}

\fb{ {\large \bf \underline{Key Question:}\\ \phantom{}\smallskip What are the properties of the first stars and galaxies?\\ }
The first structures in the Universe formed at $z>7$, at higher redshift than all currently known galaxies.  Since GRBs are brighter than other cosmological sources at high redshift and exhibit simple power-law afterglow spectra that is ideal for absorption studies, they serve as powerful tools for studying the early universe.   New facilities planned for the coming decade will be able to obtain a large sample of high-redshift GRBs.  Such a sample would constrain the nature of the first stars, galaxies, and the reionization history of the Universe.
}

\section{Introduction}

\begin{figure}
\begin{center}
{\epsfig{file=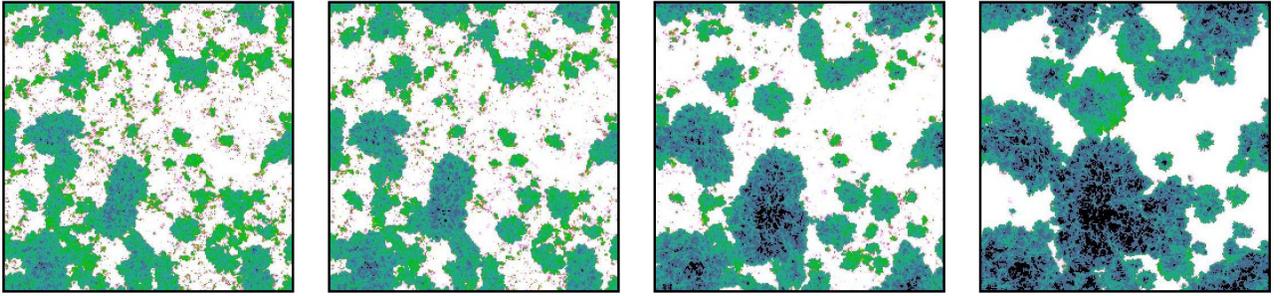, width=17.0cm}}
\end{center}
\caption{\it \small Illustration of the structure of reionization depicting slices through four simulations of reionization.  The average gas element in the IGM is half ionized in these slices (the black regions are very ionized $[x_H \sim 10^{-5}]$, the green are intermediate $[x_H \sim 10^{-3}]$, and the white are neutral $[x_H \sim 1]$), and each slice subtends $100$ comoving Mpc or $0.5$ degrees on the sky.  From left to right, the slices are from models in which progressively more massive galaxies ionize the IGM.  As apparent, understanding the nature of reionization requires a statistically representative sample of sightlines.  From \citep{mcquinn06b}. 
\label{fig:slice}}
\end{figure}

The epoch between the formation of neutral hydrogen ($z \sim 1100$) and the highest redshift quasars and GRBs that have been observed ($z \sim 6.5$) saw a number of milestones for the early universe: the first stars were formed, the first galaxies were assembled, and the first massive black holes were created.  These first objects produced the ionizing photons that ``reionized'' the hydrogen in the Universe.  Absorption-line studies of $z > 6$ quasars (QSOs) suggest that the reionization of hydrogen was very nearly complete by $z \approx 6$ (e.g. \citep{fck+06}).  Yet, based on polarization studies of the cosmic microwave background, reionization appears to have started much earlier ($z \gtrsim10$) \citep{2008arXiv0803.0586D}.  Taking these observations at face value would imply an extended reionization process that spans several hundred million years.  Perhaps a single population of sources ionized the Universe between $z \sim 10$ and $z \sim 6$. Alternatively, the Universe could have been partially ionized at $z \sim 10$ by one set of UV-emitting objects, then reionization stalled, then was fully ionized after the emergence of a new population of UV-bright sources. 
With presently little or no {\it in situ} observations, the objects responsible for reionization are envisioned theoretically
in a far-reaching array of possibilities: the likely culprits include
Pop III and II stars (the $1^{\rm st}$ and $2^{\rm nd}$ generations of stars) \citep{wl03}, non-thermal emission from mini-QSOs or supernova (SN) shocks \citep{mrv+04, oh01}, QSOs, and even decaying/annihilating particles (e.g., dark matter) \citep{mfp06}.  Figure \ref{fig:slice} illustrates four stellar reionization models from \citet{mcquinn06b}.

Identifying the timeline and sources of reionization will be a significant endeavor in the next decade, with a variety of approaches.  In addition to questions regarding reionization, astronomers aim to address several other questions related to the first stars and galaxies: 
\begin{itemize}	
	\item {\bf When is the IGM enriched with metals?  How do metals escape from galaxies?}  The intergalactic medium (IGM) appears highly enriched in metals at $z \lesssim 6$ ($\sim 10^{-3}$ Solar).  Absorption lines of \ion{C}{IV} and \ion{O}{IV} are ubiquitous in the $z<6$ Ly$\alpha$ forest, and all HI column densities in the Ly$\alpha$ forest are found to be enriched to $\sim 10^{-3}$ Solar in metals  \citep{cowie98}.  These facts suggest the IGM was enriched at early times to $~10^{-3}$ Solar.

\item {\bf Do Population III stars exists and in what environments?} Analytic studies and numerical simulations suggest that the first generation of stars were massive ($\sim 100 \; \Msun$), hot ($\sim 10^5$ K), and formed in tiny dark matter halos ($\sim 10^6~M_{\odot}$).  Observations have yet to test these predictions.

\item {\bf What are the masses and makeup of the first galaxies?}  The first galaxies were hundreds of times less massive than the Milky Way.  It is these galaxies that probably ionized the Universe.  Therefore, they must have been able to efficiently produce ionizing photons that escaped into the IGM, and some must have been able to survive potentially fatal feedback mechanisms  (such as SN and photo-heating feedback).  Current surveys capture the most massive galaxies at high redshift and none at $z>7$.

\item {\bf What are the fundamental cosmological parameters that describe the constituents and initial conditions of the Universe?} Knowledge of the reionization history will break degeneracies between  the reionization history and fundamental cosmological parameters.  For example, the constraint on the tilt of the primordial power spectrum (which reveals a lot regarding inflation) would be significantly tighter if the reionization history were known \citep{komatsu08}.  Improved knowledge of the Cosmology can significantly impact predictions for when the first galaxies formed.  

\end{itemize}
	
\noindent High-redshift gamma ray bursts (GRBs) are one of the most promising probes to investigate these interesting questions.  The afterglow flux from a GRB at a fixed time since the prompt emission is nearly independent of its redshift \citep{ciardi99, lamb01}. This makes GRB afterglows by far the brightest beacons at high redshift for hours to days after the prompt event, and their power-law afterglow spectrum is ideal for galactic and intergalactic absorption studies.   

\section{Using Gamma-Ray Bursts as Probes}
\label{sec:2}
Since the prompt hard X-ray and $\gamma$-ray emission of GRBs is not attenuated by gas and dust, 
GRBs should in principle be detectable from within the reionization
epoch.  While there are large uncertainties in modeling the rate of GRBs at high redshift, studies find that at least a few percent of Swift GRBs should originate from beyond $z = 7$ \cite{gmaz04,bl06, salvaterra08}.  The recent discoveries of $z=6.29$ and $z=6.7$ GRBs with Swift and ground-based near-infrared (NIR) followup  \citep{hnr+06,kawai05,2008arXiv0810.2314G,Fynbo08_GCN8225} confirm that GRB-progenitor stars exist in the early universe and that their GRBs are detectable with current instruments.  However, the lack of on-board near-infrared capabilities and the small ($30$ cm) optical aperture on Swift has almost certainly resulted in high-redshift events going unidentified.  Section \ref{EXIST} outlines the requirements to efficiently detect high-redshift GRBs and to study the high-redshift universe with their afterglow spectrum, and describes EXIST, a mission that is optimized for this science.

\begin{figure}
\begin{center}
{\epsfig{file=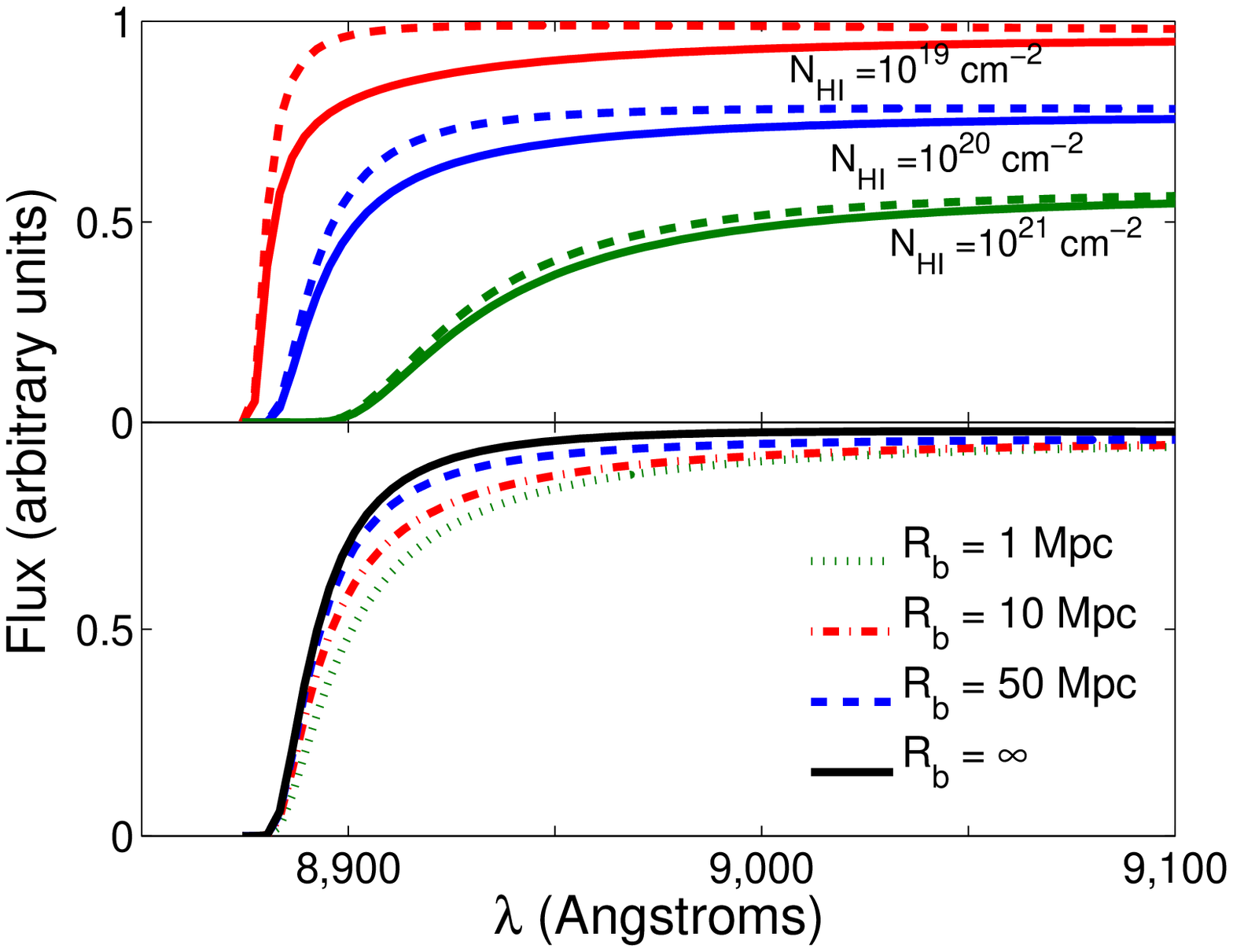, width=8.1cm}
\epsfig{file=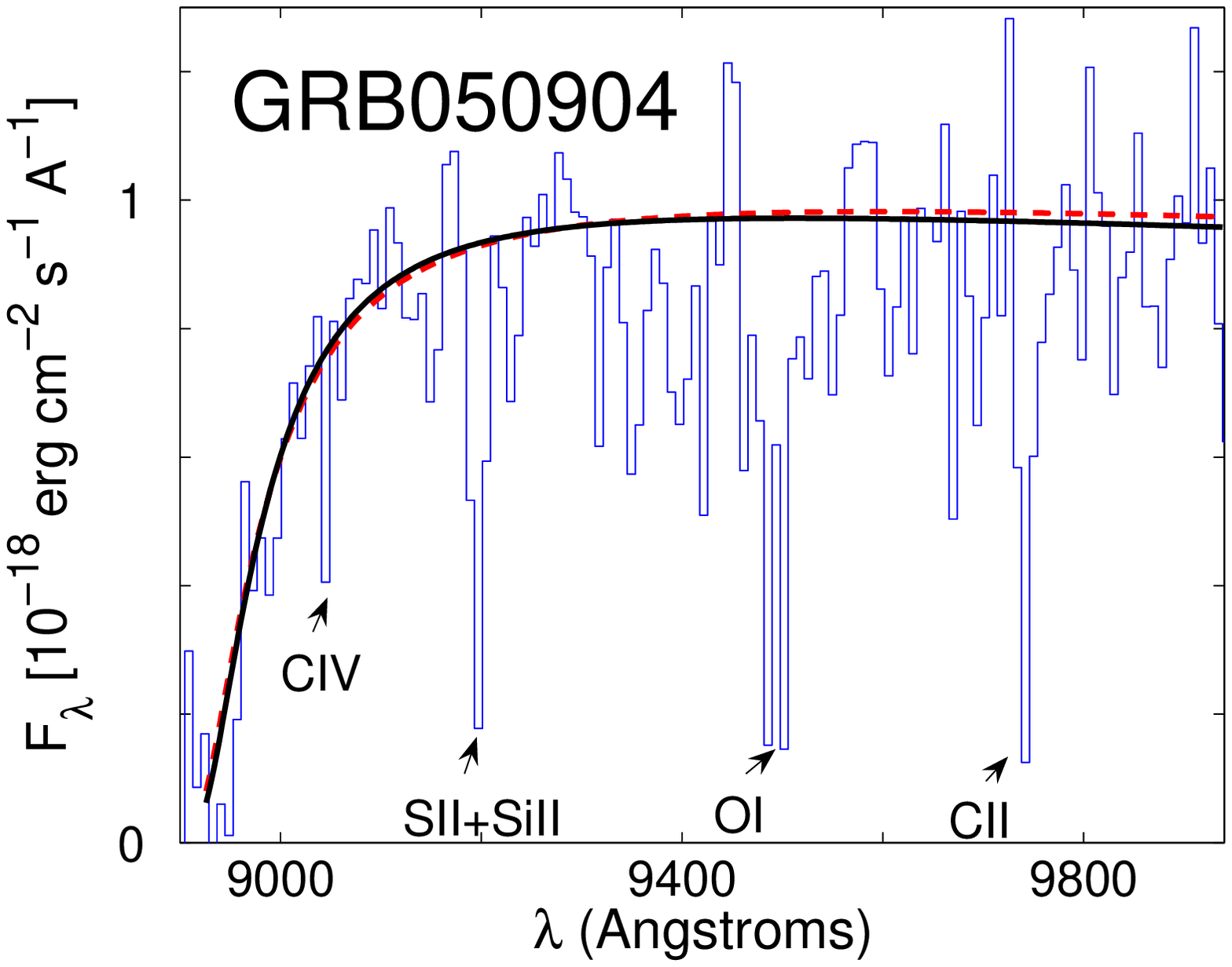, width=8.1cm}}
\end{center}
\caption{\it \footnotesize Left: Illustration of the GRB afterglow flux near the GRB-frame
Ly${\alpha}$ wavelength.  The top panel depicts the afterglow flux for
both an ionized IGM (dashed curves) and one that is half ionized
where the GRB sits in a $10$ comoving Mpc bubble (solid curves), and both models include absorption from the host galaxy with column density $N_{\rm HI}$.  Half of GRBs have $N_{\rm HI} < 10^{21}   ~ {\rm
  cm}^{-2}$. The bottom panel illustrates the
effect of the host bubble comoving size on the afterglow flux,
assuming the Universe is half ionized and $N_{\rm HI} = 10^{20} ~ {\rm
  cm}^{-2}$.  The size of the bubble has implications on how the Universe was reionized (see Fig. 1).    Right:  The afterglow spectrum of the $z = 6.3$ burst
GRB\, 050904 (blue histogram) \citep{kawai05, totani05}, as well as
two best fit spectral models:  A model that only includes absorption
from the host galaxy (solid black curve) and a model that also
includes IGM absorption (red curve).  Both models require  $N_{\rm HI}
= 10^{21.6} ~ {\rm cm}^{-2}$, which obscures the IGM contribution. 
\label{fig:dampingwing}
\label{fig:GRB050904}}
\end{figure}

The detection of a high-redshift GRB as well as high signal-to-noise followup measurements of its afterglow spectrum will enable science that is often not possible with other high-redshift probes, including\\ 

\noindent {\bf The IGM Neutral Fraction and HII Bubble Sizes during Reionization}: GRB afterglow absorption is sensitive to the structure of reionization as well as the global history and, therefore, has the potential to distinguish between different reionization models \citep{miralda98, mcquinn08, mesinger08}.  Emission within $100 \, {\rm \AA}$ of Ly$\alpha$ can be scattered in the damping wing of the Ly$\alpha$ line by intergalactic neutral hydrogen that is as far as $10$ proper Mpc from the source.  This scattering results in an extended Ly$\alpha$ absorption profile redward of the rest-frame Lyman-$\alpha$ line center.  This absorption is  sensitive to IGM neutral fractions of order unity, unlike the Lyman-$\alpha$ forest absorption (the absorption from scattering in the line center) which is saturated for neutral fractions of $10^{-4}$.  While all sources during reionization will suffer from this absorption, a bright source with a simple intrinsic spectrum is necessary in order to isolate it.  Because GRB afterglows are the brightest sources at high-z and have a simple power-law spectrum, they are much better candidates for this technique than galaxies or quasars.

The presence of host galaxy absorption complicates the
determination of the ionization state of the IGM in a GRB spectrum, but this absorption is less extended in wavelength and therefore can be separated in principle. Chen et al. \citep{chen07} found that $\sim$20--30\% of GRBs have small enough intrinsic HI columns to allow a direct measurement of the absorption from a partially neutral IGM.  The left panels in Figure \ref{fig:GRB050904} show this absorption for various bubble sizes during reionization and host-galaxy obscuring columns ($N_{\rm HI}$), assuming that the IGM is half ionized outside of the bubble.  A large sample of high-redshift GRBs is necessary to beat down the large cosmic variance between individual sight-lines and constrain the properties of reionization.  To make important qualitative statements about reionization (such as whether it is occurring and about the bubble sizes) requires at least at least a few bursts per redshift with $N_{\rm HI} \lesssim 10^{20}$ cm$^{-2}$, whereas to distinguish between the reionization models shown in Figure \ref{fig:slice} requires an even larger sample of GRBs \citep{mcquinn08}.  

The afterglow spectrum of the $z = 6.3$ burst GRB\, 050904 is shown in the right panel in Figure \ref{fig:GRB050904}.  This spectrum was taken $3.4$ days after the prompt emission with the Subaru Telescope \citep{kawai05}.  The flux would have been over a hundred times larger if it were instead taken an hour after the GRB trigger!   Unfortunately, the host galaxy absorption dominates the absorption redward of the Ly$\alpha$ line, preventing constraints on reionization \citep{totani05, mcquinn08}. \\

\noindent {\bf Constraints on Metal Enrichment History and Nature of Pop III stars}: 
Ten and thirty meter-class telescopes can measure the $z>6$ IGM and ISM
metallicity with the brightest GRB afterglows \citep{oh02}.  For fainter bursts where the metallicity
cannot be accurately measured
(since metal lines are unresolved in resolutions less
than R $\sim$ 20,000), the measurement of the relative gas columns
from absorption lines can yield
a \emph{pattern of abundances} that would reflect the enrichment history \cite{oum+06, hartmann08, campana08}. For
instance, if most Pop III stars explode as pair-instability
SN, then the ISM and IGM would be polluted with less Fe-peak elements
than if pollution occurred from ordinary SNe.  Other high-redshift objects are not bright enough at high-z to allow detailed enrichment studies.
Metal absorption lines can clearly be identified in the spectrum of the $z=6.3$ burst GRB 050904 (right panel, Fig. \ref{fig:GRB050904}, \citep{totani05}.  The CIV line is from an intervening system and not the host galaxy.).  \\

\noindent {\bf Fraction of Ionizing Photons that Escape from Galaxies}: This number, $f_{\rm esc}$, is very uncertain, and  different theories make contrary predictions for its value in high-redshift galaxies.
  Yet, ionizing photons must escape if stars ionize the Universe.  Ionizing photons can only escape if the hydrogen column density $N_{\rm HI}$ is less than $\sim 5 \times 10^{-17}$ cm$^{-2}$.  Observations of the damping wing feature in GRB afterglows (left panels, Fig. \ref{fig:GRB050904}) and absorption in the soft-Xray can derive the value of this parameter along the line of sight to a GRB.  Chen et al. \citep{chen07} measured $\langle f_{\rm esc} \rangle \approx 0.05$ with the damping wing technique using $27$ GRBs, but a larger high-redshift sample is required to measure it in, say, $z>5$ GRBs.\\     
\begin{figure}
\centerline{\psfig{file=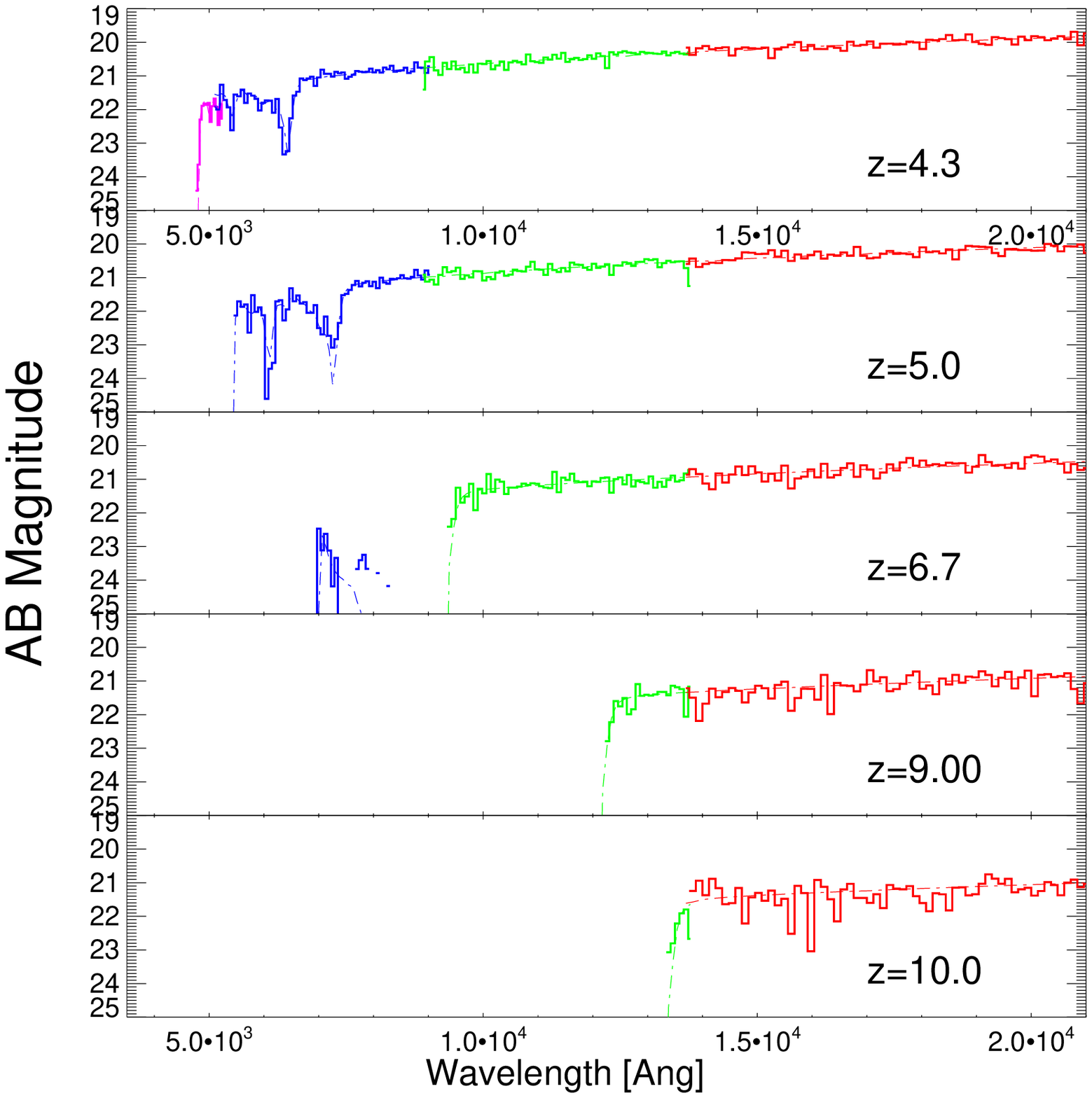, width=7cm}
\psfig{file=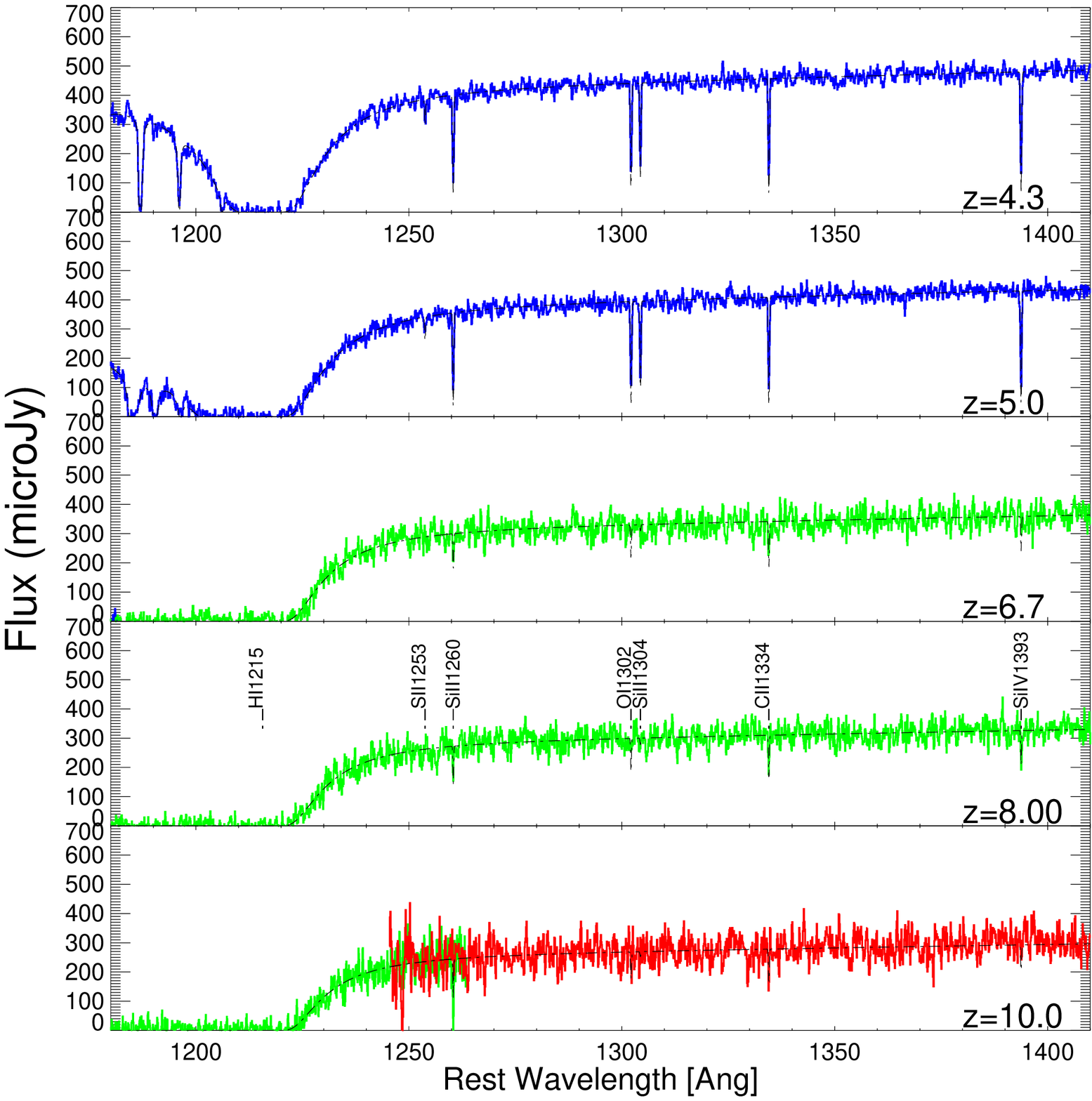, width=7cm}}
\caption{\it \footnotesize Simulations of observations of GRB afterglows using the $4$-channel optical-IR telescope on EXIST. (Left) Resolution $R=30$ grism spectrum of what would be
  one of the faintest in the Swift sample of GRBs (1.0 mag fainter
  than GRB 080913 [$z=6.7$]). (Right) Resolution $R=3000$
  cross-dispersed spectrum of a typical GRB afterglow (2.0 mag
  brighter than GRB 080913) as it would be observed.  Even with low metallicity ($Z = -4$ assumed for $z>6$), several ISM lines
  would be detected to $z \sim 10$. The colors show  simultaneous 
  spectra in the 0.3 - 0.52 $\mu$m (purple; 
upper left spectrum), 0.52 - 0.9 $\mu$m (blue), 0.9 - 1.38 $\mu$m (green), and 1.38 -
  2.2 $\mu$m (red) bands. 
\label{fig:sim}}
\end{figure}

\noindent {\bf The ISM of High-Redshift Galaxies}:  The ISM will be different at high redshift:  It should be less enriched with metals, it should be denser, there should be less dust (and the composition may be different), and magnetic fields and cosmic rays may be less important.   GRBs are the only probe that can provide high enough signal-to-noise spectra to study the $z \gtrsim 6$ ISM in detail.  Measurements of metal and molecular lines, dust attenuation, and relative velocities of different lines in GRB afterglow spectra (as well as the effect of the GRB emission on the evolution of these lines) will allow studies of the ISM properties in high-z galaxies.  These techniques are well established for Swift GRBs in the optical/NIR \cite{prochaska08} and in the X-ray \citep{campana07, campana08}.  Simultaneous measurements in the NIR and X-ray bands with EXIST (see Section 3) will provide a handle on dust properties at high redshift \citep{stratta07}. \\

\noindent {\bf Better Measurements of the High-Redshift Ly$\alpha$ Forest}:
  Unexplained, large opacity fluctuations are present in observed spectra, even on
  $\sim 50$ comoving Mpc scales, and the average opacity is increasing
  rapidly with increasing redshift  \citep{fan06}.  Many have
  interpreted these trends as evidence for reionization, but this
  interpretation remains controversial \citep{lidz06a, brs+07}.   GRBs are the best tool to measure the Ly$\alpha$ forest at $z>6$.  At high-z, where there is little transmission in the forest, the continuum emission can be more cleanly removed from GRB afterglow spectra than from quasar spectra, enabling more detailed studies of the absorption. 
    Furthermore, the abundance of bright quasars is
  falling off more rapidly with redshift than that of GRBs.    A sample of Ly$\alpha$ forest spectra from high-z GRBs may even provide competitive constraints on the neutral fraction of the IGM \citep{gallerani08}.\\

\noindent {\bf Signposts to the First Galaxies}:  GRB-selected galaxies are close to holding the record for the highest redshift galaxy, and, because GRBs are brighter than galaxies and quasars at high redshift, it is likely that  the next record holder will be located with a GRB.   In addition,  GRB-selected galaxies will provide a population of spectroscopically confirmed high-redshift galaxies for  next-generation telescopes like ALMA, GMT, JWST, and TMT to study (perhaps the only known high-z sources that will fall outside the narrow deep fields of these telescopes).   GRB-selected galaxies are expected to trace star formation, and, therefore, should be \emph{more representative} than galaxies that are selected by other means. \\

\noindent {\bf Constraints on the Nature of Dark Matter}:  There are few constraints on the spectrum of primordial fluctuations on mass scales below $10^{10} \, M_\odot$, and, while the nominal assumption is that a single power-law index measured at high mass continues to extremely low masses, there are processes that suppress small-scale structure (and perhaps reconcile the missing dwarf galaxy problem exigent in Lambda CDM).  Warm dark matter models (via particle free streaming) and reionization itself (via heating the gas) could wash out fluctuations on dwarf-galaxy scales.  This could drastically change the number of collapsed halos at high redshift, the homes of GRB progenitors.  The detection of a single GRB at $z = 10$ would rule out models in which a $2$ keV thermal relic particle is the dark matter, and the constraints are even stronger if a GRB is observed at higher redshift  \citep{mesinger05}.\\

\noindent {\bf The Star Formation History}:  The redshift distribution of a large sample of GRBs provides an independent handle on the star formation history \citep{hartmann08}.  It is unclear how well the GRB rate traces the star formation rate (SFR).  Kistler et al. \citep{kistler07} have argued that there are too many GRBs at high-z for it to be a direct tracer, while Faucher-Gigu{\`e}re et al. \citep{faucher08} suggested that GRBs may trace the SFR at high-z better than other more direct measures of star formation.  Since GRBs can be observed at higher redshift than all other cosmological objects, high-z GRB surveys may be the only means to directly verify whether stars exist at $z \gtrsim 12$.  EXIST can measure redshift via Lyman-breaks up to $z \approx 18$.

\section{An Integrated Space-based Approach to the Study of Reionization}
\label{EXIST}


Enabling such compelling science requires sensitive, wide-field detection of GRBs in the hard-X-ray and soft $\gamma$-ray regimes to find
an appreciable number of faint and rare high-redshift events. As Swift
has helped demonstrate, the precise localization of the
events is most robust ($>95\%$ success with Swift) when rapid on-board X-ray followup can be performed.  Ground-based followup is
simply not a reliable venue for afterglow discovery and
localization.  Swift has also demonstrated the utility of on-board optical/UV observations, with a modest
detection rate of afterglows in these bands with its 30 cm
telescope. One clear lesson from Swift is that larger apertures for afterglow discovery of the fainter highest redshift
events.  Furthermore, most afterglow photons with rest-frame Ly$\alpha$
($\lambda_{\rm observed} = 1216 \, [1 + z] \, {\rm \AA} $) will be
absorbed by the IGM.   To insure proper followup of events at $z > 7$ (which are invisible in the optical), there should be on-board spectroscopic capabilities in the NIR.   There are two proposed missions that aim to locate high-z GRBs with on-board NIR capabilities: a Small Explorer-class mission, JANUS (which aims for a GRB discovery rate comparable to that of Swift), and a Medium Class Mission, EXIST.  

EXIST (the Energetic X-ray Imaging Survey Telescope) currently in pre-Phase A planning as an  Astrophysics Strategic Mission
  Concept Study,  includes a broad-band (5--600 keV) wide-field
  ($\sim90^{\circ}$) sensitive imager ($< 20"$).  This imager is $\sim 7\times$ more sensitive than Swift over the same band, but
  extends to both significantly lower and higher energies, and this allows EXIST to
  detect 500 -- 700 GRBs per year ($5-7 \times$ Swift).   The prompt emission from high-z bursts is fainter and has lower peak energies, and studies find that Swift will not trigger on many of these events \citep{salvaterra08}.  Even for bright high-z bursts Swift does detect, its lack of NIR camera prevents it from measuring the redshift.   EXIST should detect and measure redshifts for many more $z>6$ GRBs than the $2$ that Swift (with the aid of ground-based followup) has detected in $4$ yr.   Forecasts are that EXIST would obtain $50-300$ GRBs at $z > 6$ in $5$ yr of operation \citep{salvaterra08}, enabling detailed studies of the structure of reionization.  A Soft X-ray Imager/spectrometer (SXI) and co-aligned $1.1$m optical-IR 
telescope (IRT) for both imaging and spectroscopy will point to 
GRB positions within $\sim 2$ min.  The afterglow will be identified by either
  dropouts or variability in the simultaneous 4-band imaging (AB = 25) in the $<3$ arc-second X-ray localization error circle and
  surrounding $4$ arc-minute field  (see Fig. 3 for the four simultaneous bands).  For more than 95\% of events to $z = 10$, EXIST should measure the redshift to $\Delta z/z <
  0.01$. For sources brighter than $H=20$ mag (AB) a high-resolution
  spectrum ($R=3000$) would be obtained, the resolution needed to study reionization \citep{mcquinn08, gallerani08}. 
  
  Nominally half of EXIST events at $z=8$ should be bright
  enough for on-board high-resolution  ($R = 3000$) spectroscopy, and, for $\age 50$ sightlines, it will be able to
  directly measure Lyman-$\alpha$ damping wings to constrain locally
reionization.   Figure 3 shows a mock $R=30$ ($3000$) spectrum with EXIST for a faint (average) luminosity burst. However,
  irrespective of brightness distribution at high redshift, EXIST will
  provide precise locations in the sky and in redshift space to high-z events and galaxies. This gives EXIST an important
    synergistic  connection to large aperture facilities (e.g.,  GMT and TMT) and
  JWST, which would be prime facilities for followup studies of
  the host galaxies and neighboring structures. The advantages of 
low backgrounds from being in space (particularly in the NIR) as well
as the $\ge 5$ magnitude brightness advantage of being
on-source in minutes rather than hours, make the EXIST-IRT an
essential tool in mapping out the first galaxies and patchy reionization.



\bibliographystyle{apj1c}

\begin{multicols}{3}
\begin{spacing}{.9}
\begin{scriptsize}

\end{scriptsize}
\end{spacing}
\end{multicols}

\end{document}